\documentclass[dvips,12pt] {article}
\usepackage{epsfig}
\usepackage{rotate}
\usepackage{a4wide}
\usepackage{amssymb}
\usepackage{amsmath}

\def\lsim{\mathrel{\rlap{\raise 2.5pt \hbox{$<$}}\lower 2.5pt
\hbox{$\sim$}}}
\renewcommand{\thefootnote}{\fnsymbol{footnote}}

\def\air{\vphantom{\Big|}}

\begin{document}

\thispagestyle{empty}

\begin{center}
{\bf \Large Neutrinoless double beta decay, solar neutrinos \\[6pt]
and mass scales}
\end{center}
\vspace{0.5cm}

\begin{center}
Per Osland\footnote[1]{E-mail: per.osland@fi.uib.no} and
Geir Vigdel\footnote[4]{E-mail: geir.vigdel@fi.uib.no}
\end{center}
\vspace{1cm}

\begin{center}
Department of Physics, University of Bergen,
     Allegt.\ 55, N-5007 Bergen, Norway
\end{center}
\vspace*{1.5cm}

\begin{abstract}
We obtain bounds for the neutrino masses by combining atmospheric
and solar neutrino data with the phenomenology of neutrinoless double
beta decay where hypothetical values of $|\langle m\rangle|$ are envisaged
from future $0\nu\beta\beta$-experiments.
Different solutions for the solar neutrino data are considered.
For the Large-Mixing-Angle solution, a bound
$|\langle m\rangle|\leq0.01~\text{eV}$ would strongly disfavour an inverted
hierarchy of the neutrino masses.
\end{abstract}
\clearpage
\renewcommand{\thefootnote}{\arabic{footnote}}
\section{Introduction}
While the two well-known neutrino anomalies, the atmospheric and solar
neutrino problems, can be explained in terms of mixing angles and
mass-squared differences, the absolute neutrino masses remain largely
unknown. There are theoretical models which relate the mixing angles to the
masses, and thereby for neutrino oscillation data provide fits in terms of
the masses.  Since we in this article are trying to deduce mass-bounds
based on as few assumptions as possible, such models will not be considered.

There are several (more or less direct) methods for measuring the absolute
neutrino masses, but so far none of them has given unequivocal lower bounds.
The present most stringent bound from {\em direct} measurements of neutrino
masses is derived from measuring the end-point energy of electrons in tritium
decay. The bound is $m_{\nu_e} < 2.5~\mbox{eV}$ \cite{Lobashev:2001uu}, and
since this is well above the results from the other methods, it will not be
discussed here.  The relevance for the neutrino to be part of the Dark Matter
has declined in the last couple of decades.  Once, the individual neutrino
mass was assumed to be $m_j \sim 10 ~\mbox{eV}$, but the current upper limit
is $m_j \simeq 1.8 ~\mbox{eV}$ \cite{Croft:1999mm}, based on cosmological
models of galaxy structure formation and the cosmic microwave background
radiation.  Moreover, we have the fascinating proposal that the cosmic ray
spectrum beyond the GZK cutoff \cite{Bhattacharjee:2000qc} could be due to
Z-bursts \cite{Pas:2001nd,Weiler:1982qy} induced by ultra-high-energy
neutrinos interacting with relic cosmic neutrinos, and that a study of this
spectrum could provide bounds on the neutrino masses.

Here, we will mostly be concerned with neutrinoless double beta decay, which
would occur only for Majorana neutrinos.  Some relevant numbers derived from
the different methods are given in Table \ref{Tab:masses}.

\begin{table}[ht]
\begin{center}
\begin{tabular}{|cc|c|c|}\hline
& Mass bounds [\mbox{eV}]&Future bounds [eV] & Comments \\ \hline
\makebox[3.5cm]{Tritium}\vline\air & $m_{\nu_e} < 2.5$ \cite{Lobashev:2001uu}&
$m_{\nu_e} < 0.3$ \cite{Planned_mnue} & \\
\makebox[3.5cm]{Cosmological}\vline\air & $\sum_j m_j \lesssim 5.5$
\cite{Croft:1999mm} &$\sum_j m_j \lesssim 0.3$ \cite{Croft:1999mm} &  \\
\makebox[3.5cm]{Z-burst}\vline\air & $0.1 \lesssim m_{3} \lesssim 1$
\cite{Pas:2001nd}&& Speculative \\ \makebox[3.5cm]{$0\nu\beta\beta$}\vline\air
& $|\langle m\rangle| < 0.26$ \cite{Klapdor-Kleingrothaus:2001dg} &$|\langle
m\rangle|<0.01- 0.001$
\cite{Klapdor-Kleingrothaus:2001dh}& Majorana\\ \hline
\end{tabular}
\end{center}
\caption{Four methods to explore small neutrino masses. The first and last
measure linear combinations of the masses; the Z-burst model is
sensitive to the heaviest mass state. The parameters are defined below.}
\label{Tab:masses}
\end{table}

As is well known, the solar and atmospheric neutrino oscillation phenomena
depend on the masses and mixings in a way which is rather different from the
corresponding dependency in neutrinoless double beta decay. The relationship
between these phenomena has been studied in several articles, e.g.\
\cite{K-KPS_Bilenky}.  Our aim in this paper is to further elucidate this
connection, and discuss how a possible future signal from a
$0\nu\beta\beta$-experiment, combined with increased precision in the
oscillation data, can provide constraints on the {\em absolute} values of the
neutrino masses.
\section{Oscillation parameters and \boldmath{$0\nu\beta\beta$}}
We assume the mass-squared differences, $\Delta m_{kj}^2=m_k^2-m_j^2$, are
fixed by the values relevant for the atmospheric and solar neutrino data.  In
a plane spanned by mass-squared difference and mixing angle, there are four
main regions which provide good fits to the solar neutrino problem, see Table
\ref{Tab:ranges}.
\begin{table}[h]
\begin{center}
\begin{tabular}{|rcc|cc|c|}\hline
& & \hspace*{-2.3cm}$\underline{\tan^2\theta}$ & &
\hspace*{-2cm}$\underline{\Delta m^2\ [\mbox{eV}^2]}\air$& g.o.f.\\
&Min. & Max. & Min. & Max. &  \\ \hline
ATM\hspace*{2mm}\vline\air& 0.3 & 1 & $3\times 10^{-4}$ &
$9\times 10^{-3}$ & 54\% \\
LMA\hspace*{2mm}\vline\air & 0.2 & 2& $2\times10^{-5}$ & $4\times 10^{-4}$&
59\% \\
LOW\hspace*{2mm}\vline\air & 0.4 & 3 & $2\times10^{-9}$ &
$3\times 10^{-7}$& 45\% \\
VO\hspace*{2mm}\vline\air & 0.2 & 5 & $10^{-10}$ & $10^{-9}$ & 42\% \\
SMA\hspace*{2mm}\vline\air & $2\times10^{-4}$ & $9\times10^{-4}$ &
$4\times10^{-6}$ & $1\times 10^{-5}$& 19\% \\ \hline
\end{tabular}
\end{center}
\caption{The ranges of the observed neutrino parameters within $\sim$ 99\%
C.L., as read off from figures (their exact values are not very important for
the present discussion). Data for the Large-Mixing-Angle, LOW,
Vacuum-Oscillation, and Small-Mixing-Angle solutions are from
\cite{Bahcall:2001zu}; and the data for the ATMospheric neutrino observation
is from \cite{Fukuda:1998mi}.}
\label{Tab:ranges}
\end{table}
When both
the atmospheric and solar mass-squared differences are stretched to their
limits, they can have the same value; but in such a case the fit between
theory and data is quite poor. The best fits are obtained with $\Delta m_{\rm
atm}^2 \simeq 3\times 10^{-3}~\mbox{eV}^2$ and $\Delta m_\odot^2 \simeq
4\times 10^{-5}~\mbox{eV}^2$.  Thus we consider two possible arrangements of
the relative mass-squared differences,
\begin{equation}\label{Possibilities}
\mbox{Spectrum 1:}\ \Delta m_{21}^2 = \Delta m_\odot^2,\qquad
\mbox{Spectrum 2:}\ \Delta m_{21}^2 =\Delta m_{\rm atm}^2,
\end{equation}
where, for both spectra, the mass states are denoted such that their
respective masses satisfy $m_1 < m_2 < m_3$.
In both cases we necessarily have $\Delta m_{31}^2 \simeq \Delta m_{\rm
atm}^2$.  Measured in terms of masses, two of the states will be close
together and one more apart.  A good fit requires a weak coupling between the
electron neutrino and the lone mass state which is responsible for the
largest mass-squared difference, $\Delta m_{\rm atm}^2$.

In much of the literature one assumes Spectrum 1 and $\nu_\alpha = \sum_j
U_{\alpha j} \nu_j$ as the connection between flavour and mass states.  Then,
the coupling between the electron neutrino and the most isolated mass state
will naturally be denoted as $U_{e3}$.  Because of this historical fact, we
will use $U_{e3}$ as the strength between the electron neutrino and the lone
mass state also for Spectrum 2.  Accordingly, shifting from one spectrum to
the other corresponds to a cyclic permutation.  With two $\Delta m_{kj}^2$
fixed, all three neutrino masses can be expressed in terms of {\em one} mass,
which we take to be the lightest one, $m_1$.  There are three main hierarchy
types:
\begin{eqnarray}\label{hierarchy}
\mbox{Normal hierarchy:}\quad m_1 &\ll&
\sqrt{\Delta m_\odot^2}=\sqrt{\Delta m_{21}^2}
\ll \sqrt{\Delta m_{\rm atm}^2}=\sqrt{\Delta m_{32}^2},\nonumber \\ &&
\mbox{gives } m_1 \ll m_2 \ll m_3. \nonumber \\
\mbox{Degenerated masses:}\quad m_1 &\gg& \sqrt{\Delta m_{\rm atm}^2}=
\sqrt{\Delta m_{32}^2} \gg \sqrt{\Delta m_\odot^2}=\sqrt{\Delta m_{21}^2},
\nonumber \\ &&\mbox{gives } m_1 \simeq m_2 \simeq m_3. \nonumber \\
\mbox{Inverted hierarchy:}\quad m_1 &\ll& \sqrt{\Delta m_\odot^2}=\sqrt{\Delta
m_{32}^2} \ll \sqrt{\Delta m_{\rm atm}^2}=\sqrt{\Delta m_{21}^2} \nonumber \\
&& \mbox{gives }m_1 \ll m_2 \lesssim m_3.
\end{eqnarray}
These are the ``extreme'' cases, of course $m_1$ can be close (or equal) to
the square root of some mass-squared difference.  It is convenient to
introduce a quantitative criterion for when the neutrinos can be called
degenerated.  If we define the criterion for degeneracy to be $m_1/m_3 >
0.99$, then, with $\Delta m_{31}^2 = 3.3 \times 10^{-3}~\mbox{eV}^2$,
degeneracy is achieved when $m_1 \gtrsim 0.4 ~\mbox{eV}$. If the criterion
reads $m_1/m_3 > 0.90$, then the neutrinos are degenerated for $m_1 \gtrsim
0.1 ~\mbox{eV}$.
As we will see, among the methods listed in Table
\ref{Tab:masses}, it is only $0\nu\beta\beta$-experiments that for
non-degenerated masses can give a positive signal in the foreseeable future.

For some unstable elements normal beta disintegration is forbidden by
energetic reasons, but {\em double} beta decay may be allowed. This is a
higher order process in which two nucleons decay at the same time, most of
these reactions are of the form
\begin{equation}
^A_ZX \rightarrow {_{Z+2}^{\phantom{z+}A}}X + 2e^- + 2\bar{\nu}_e .
\end{equation}
If the electron neutrino emitted from a nucleon is a Majorana particle with
non-zero mass, then it has a non-zero probability to be right-handed and
thereby it can be absorbed as an antineutrino by a nucleon of the same type
as the one from which it originated.  Thus, the final state of this reaction
contains no neutrino,
\begin{equation}\label{0nubetabeta}
^A_ZX \rightarrow {_{Z+2}^{\phantom{z+}A}}X + 2e^-.
\end{equation}
There are several other ways for $0\nu\beta\beta$ to occur. In this
article we consider as small an extension of the Standard Model as possible,
therefore we assume that only left-handed charged currents are involved and
that the above mechanism takes place by the exchange of a light Majorana
neutrino\footnote{This could be due to a Higgs triplet or heavy Majorana
partners \cite{Chikashige:1981ui}.}.

The rate for the process (\ref{0nubetabeta}) depends on the $M_{ee}$ element
(which we hereafter denote as $\langle m\rangle$) of the mass matrix
\begin{equation}\label{def:M}
M = U^*DU^\dag,
\end{equation}
where $D$ is a diagonal matrix whose entries are the neutrino mass
eigenvalues. Depending on whether we assume Spectrum 1 or 2 (see
Eq.~(\ref{Possibilities})) $D$ will respectively be\footnote{Note that
Spectrum 2 is obtained by a cyclic permutation.}
\begin{equation}\label{Def:D}
D={\rm diag}(m_1, m_2, m_3)\quad {\rm or}\quad D={\rm diag}(m_2, m_3, m_1).
\end{equation}

To allow for the possibility of neutrinoless double beta decay, we have to
assume the neutrinos are of Majorana type. Then the mixing matrix can be
expressed as
\begin{equation}
U=\left[\begin{array}{ccc}
U_{e1} & U_{e2} & U_{e3} \\U_{\mu 1} & U_{\mu 2} & U_{\mu 3} \\
U_{\tau 1} & U_{\tau 2} & U_{\tau 3}
\end{array}\right]
\left[\begin{array}{ccc}
1 &0 & 0\\0 & e^{-i\alpha_1/2} & 0\\ 0& 0 & e^{-i\alpha_2/2}
\end{array}\right],
\end{equation}
where $\alpha_1$ and $\alpha_2$ are Majorana phases, their ranges are $0\le
\alpha_1,\alpha_2 < 2\pi$.  The ``universal'' phase is included in the left
mixing matrix. This phase can be rotated to an arbitrary $2 \times 2$ sub
matrix of $U$, and because the observable parameter in $0\nu\beta\beta$
contains mixing elements only from the first row of $U$ (see
Eqs.~(\ref{spect1}) and (\ref{spect2})), this phase is of no physical
consequence for this kind of phenomenon.

For the two spectra we get for the
electron-neutrino state and the $\langle m\rangle$-element:
\begin{eqnarray}\label{spect1}
\mbox{Spectrum 1:}&& \nonumber \\
|\nu_e\rangle&=&U_{e1}|\nu_1\rangle + U_{e2}|\nu_2\rangle
+ U_{e3}|\nu_3\rangle, \nonumber \\
\langle m\rangle&=& U_{e1}^2m_1 + U_{e2}^2m_2e^{i\alpha_1}
 + U_{e3}^2m_3e^{i\alpha_2};
\end{eqnarray}
\vspace*{-10mm}
\begin{eqnarray}\label{spect2}
\mbox{Spectrum 2:}&& \nonumber \\
|\nu_e\rangle&=&U_{e1}|\nu_2\rangle + U_{e2}|\nu_3\rangle
+ U_{e3}|\nu_1\rangle, \nonumber \\
\langle m\rangle&=& U_{e1}^2m_2 + U_{e2}^2m_3e^{i\alpha_1}
 + U_{e3}^2m_1e^{i\alpha_2}.
\end{eqnarray}
We will focus somewhat more on Spectrum 1 than the other because the first
one seems more natural, and Spectrum 2 is disfavoured for SN 1987A-neutrinos
\cite{Lunardini:2001sw}.
We note that, in contrast to neutrino oscillations, the Majorana phases have
to be accounted for in analysing results of $0\nu\beta\beta$-experiments.
Further, we see that $\langle m\rangle$ is $CP$ invariant for $\alpha_1,
\alpha_2=0$ or $\pi$, and it is useful to note that
\begin{equation}\label{begrMee}
|\langle m\rangle| \le m_3.
\end{equation}
If the future brings not only an upper bound, but a definite value for
$|\langle m\rangle|$, then Eq.~(\ref{begrMee}) yields a {\em lower} bound on
the heaviest mass.

At present, the strongest bound on $|\langle m\rangle|$ from
$0\nu\beta\beta$-measurements is \cite{Klapdor-Kleingrothaus:2001dg}
\begin{equation}\label{bounds}
|\langle m\rangle| < 0.26~\mbox{eV}\mbox{ at 68\% C.L.} \quad{\rm and}\quad
|\langle m\rangle| < 0.34~\mbox{eV}\mbox{ at 90\% C.L.}
\end{equation}
It should be noted that the exact values of the above limits depend on the
nuclear matrix elements, which have a considerably uncertainty
\cite{Vogel:2000vc}. However, for our phenomenological study that will not be
taken account of.

For neutrinoless double beta decay the difference between the two spectra
diminishes as the degree of degeneracy increases. Therefore, in deriving mass
bounds from today's upper bound on $|\langle m\rangle|$, we get essentially
the same result whether we assume Spectrum 1 or 2 because the large masses
required correspond to near-degenerated mass states.  However, as we shall
see, with the foreseen reach of GENIUS \cite{Klapdor-Kleingrothaus:2001dh},
we get spectrum-dependent bounds on the individual masses.
\section{Limiting cases}
In order to develop some intuition for the expressions in Eqs.~(\ref{spect1})
and (\ref{spect2}), we will study two realistic limiting cases.  It is
generally believed that the neutrinos are quite light, and there is no
compelling theory which imposes a lower bound for the lightest mass state.
Therefore, one limit of interest is $m_1 = 0$.  The second limiting case is
small $U_{e3}$, as is indicated by several experiments.  In either of these
two limits, the expressions for $\langle m\rangle$ reduce to two terms, where
its smallest and largest value (for given values of the masses and the
mixing) are obtained with the remaining phase equal to $\pi$ and 0,
respectively.

Some of the figures in this article will be expressed in terms of mixing
angles.  Our convention for these angles is the one advocated by the Particle
Data Group \cite{Groom:2000in},
\begin{equation}\label{Uelements}
U_{e1}=\cos\theta_{12}\cos\theta_{13},\quad
U_{e2}=\sin\theta_{12}\cos\theta_{13},\quad U_{e3}=\sin\theta_{13}.
\end{equation}

\subsection{{\bf Negligible}\ {\boldmath $m_1$}}
When $m_1$ is small, Eqs.~(\ref{spect1}) and (\ref{spect2}) can be
approximated as
\begin{alignat}{2}
|\langle m\rangle| &\simeq |U_{e2}^2\sqrt{\Delta m_\odot^2} + \
U_{e3}^2\sqrt{\Delta m_{\rm atm}^2}\,e^{i\alpha_1}|,
\quad & \mbox{Spectrum 1}, & \\[2mm]
|\langle m\rangle| &\simeq
\sqrt{\Delta m_{\rm atm}^2}|U_{e1}^2+U_{e2}^2\,e^{i\alpha_1}|,
& \mbox{Spectrum 2.}
\end{alignat}
Under this assumption of negligible $m_1$, we can of course not have
degenerated neutrinos, but Spectrum 2 would be an example of inverted
hierarchy.  The ranges of the effective Majorana mass for the SMA solution
are $0 \lesssim |\langle m\rangle| \lesssim 0.003 ~\mbox{eV}$ (Spectrum 1)
and $0.02 \lesssim |\langle m\rangle| \lesssim 0.08 ~\mbox{eV}$ (Spectrum 2).
Correspondingly for the LMA region, $0 \lesssim |\langle m\rangle| \lesssim
0.005 ~\mbox{eV}$, (Spectrum 1) and $0.003~\mbox{eV} \lesssim |\langle
m\rangle| \lesssim 0.08 ~\mbox{eV}$ (Spectrum 2), and for the LOW
solution: $0 \lesssim |\langle m\rangle| \lesssim 0.003~\mbox{eV}$
(Spectrum 1) and $0 \lesssim |\langle m\rangle| \lesssim 0.08 ~\mbox{eV}$
(Spectrum 2).  These values are well below current limits,
but the Spectrum 2 (and perhaps Spectrum 1) values can
be explored with the coming GENIUS experiment
\cite{Klapdor-Kleingrothaus:2001dh}.

\subsection{{\boldmath Small $U_{e3}$}}
One important difference between this and the former ($m_1=0$) limit, is that
the present one is compatible with all hierarchy types in
Eq.~(\ref{hierarchy}).  It is well-known that both the atmospheric and solar
neutrino data give best fits to the neutrino oscillation hypothesis when
$U_{e3} \ll 1$.  A small value for $U_{e3}$ is also strongly suggested by the
CHOOZ data \cite{Apollonio:1999ae}.  Here we will study both the ``exact''
case of $U_{e3}=0$ and the case $U_{e3}=0.2$.  The latter value is motivated
by the CHOOZ experiment, which implies (to 90\% C.L.)  $U_{e3} \lesssim 0.2$
for $\Delta m_{\rm atm}^2 > 3\times 10^{-3}~\mbox{eV}^2$.  Whether or not
this mixing element is negligible, could be determined at a neutrino factory
which would either establish a definite value for $U_{e3}$, or lower the
upper bound to $U_{e3} \lesssim 0.015$ \cite{Albright:2000xi}.  If such a low
bound should be established, there could not be more than one effective
Majorana phase.

For Spectrum 1 and negligible $U_{e3}$, the effective Majorana mass in
the SMA region can be approximated as $|\langle m\rangle|\simeq m_1$ ($\simeq
m_2$ for Spectrum 2).  If we assume Majorana neutrinos and the SMA solution,
the present upper limit on $|\langle m\rangle|$ implies $\sum_j m_j \lesssim
0.78 ~\mbox{eV}$.  The degree of degeneracy in this case is bounded by
$m_1/m_3 < 0.98$.

\begin{figure}[ht]
\begin{center}{\epsfysize=7cm
{\mbox{\epsffile{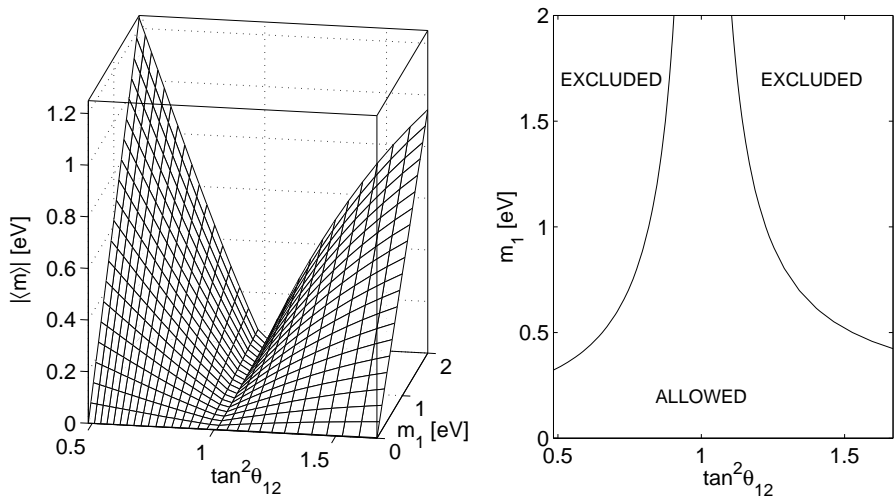}}}}
\end{center}
\caption{Left panel: $|\langle m\rangle|$ from Eq.~(\ref{apprUe3}) with
$\alpha_1 = \pi$ and $\Delta m_{21}^2 = 10^{-5}~\mbox{eV}^2$. Right panel:
Contour for today's upper limit $|\langle m\rangle|=0.26 ~\mbox{eV}$. Because
of the large value assumed for $|\langle m\rangle|$, the contour is
practically the same for Spectrum 1 and 2.}
\label{Fig:m1-Ue1-M}
\end{figure}

When $U_{e3}=0$ we get for Spectrum 1 \begin{eqnarray}\label{apprUe3}
|\langle m\rangle|&=&|U_{e1}^2m_1+U_{e2}^2m_2e^{i\alpha_1}| \nonumber \\
&=&|U_{e1}^2m_1 +e^{i\alpha_1}(1-U_{e1}^2)\sqrt{m_1^2+\Delta m_{21}^2}|.
\end{eqnarray} In Fig.~\ref{Fig:m1-Ue1-M} we display the relation between
$m_1$ and $U_{e1}$ according to Eq.~(\ref{apprUe3}).  The range of
$m_1$-values in this figure is deduced from the cosmological limit in Table
\ref{Tab:masses}.
Note that the maximal allowed $m_1$-value for the $|\langle
m\rangle|$-contour decreases as the mixing decreases.

\begin{figure}[ht]
\begin{center}{\mbox{\epsffile{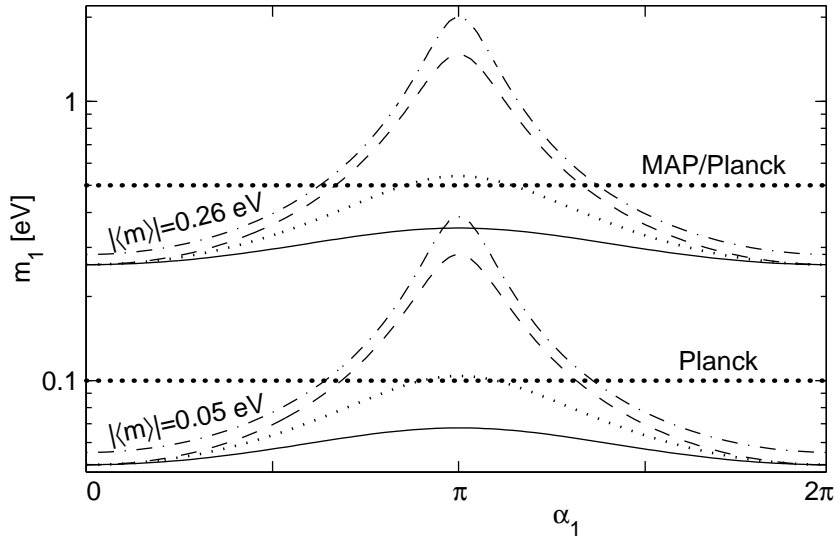}}}
\end{center}
\caption{Two sets of $\alpha_1$ and $m_1$ projections, for $|\langle
m\rangle|=0.26~\mbox{eV}$ and 0.05~eV.  The solid, dotted and dashed curves
represent $\tan^2\theta_{12}=0.15$, 0.35 and 0.7, respectively. Dash-dotted
curve: $\tan^2\theta_{12}=0.7$, $U_{e3}=0.2$, $\alpha_2=\pi$.  The lowest
horizontal dotted line shows the (most optimistic) mass bound derived from
the Planck space probe. The other horizontal dotted line is well within reach
for both spacecrafts.}
\label{Fig:alpham1}
\end{figure}

When we include a non-zero $U_{e3}$ in the $\langle m\rangle$-expressions, we
set the corresponding phase factor equal to $-1$, i.e., $\alpha_2=\pi$ in
Eqs.~(\ref{spect1}) and (\ref{spect2}). For the relevant bound on $|\langle
m\rangle|$ this choice gives the highest allowed mass values.
Fig.~\ref{Fig:alpham1} shows, for Spectrum 1, $m_1$ as
a function of the $CP$-parameter $\alpha_1$ for two values of $|\langle
m\rangle|$, namely 0.26 eV (current limit) and 0.05 eV (within the
sensitivity of GENIUS). The highest and lowest mixing in this figure
corresponds to the highest and lowest mixing allowed by the LMA region (95\%
C.L.) for two generations. As we see, the variation of $m_1$ with the phases,
depends on how strong the mixing is.  When $U_{e3}=0$, the highest possible
mass value is $m_1\simeq 1.5~\mbox{eV}$. With $U_{e3}=0.2$ the highest value
is $m_1\simeq 2.0~\mbox{eV}$, which for three neutrino generations
corresponds to $\sum_j m_j \simeq 6 ~\mbox{eV}$.  This is close to the
current upper bound from cosmological observations, given in
Table~\ref{Tab:masses}.  Results from the space probes MAP (under way) and
Planck (launch in 2007) can lead to sensitivities of $\sum m_j \simeq 0.5
~\mbox{eV}$ and $\sum m_j \simeq 0.3 ~\mbox{eV}$, respectively
\cite{Croft:1999mm}.

\section{Combination of data}
\begin{figure}[t]
\begin{center}{\mbox{\epsffile{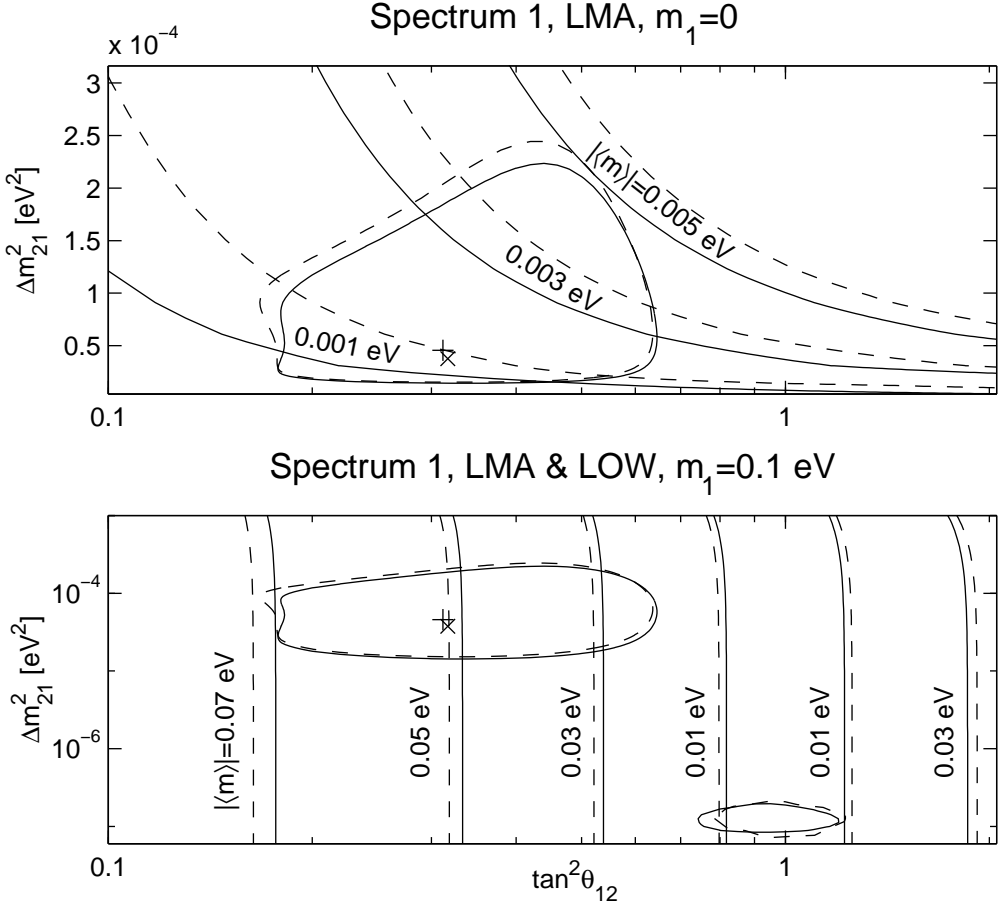}}}
\end{center}
\vspace*{-8mm}
\caption{The closed contours give the LMA and LOW (lower right) regions
allowed to 95\% C.L.  Also shown are pairs of contours of $|\langle
m\rangle|$ for $m_1=0$ (upper part) and $m_1=0.1~\mbox{eV}$ (lower part).
Solid and dashed curves represent $U_{e3}=0$ and 0.1, respectively.  For
these $U_{e3}$-values ``$\times$'' and ``+'' mark the best-fit points in the
LMA region. We set $\alpha_1 = \alpha_2= \pi$.}
\label{Fig:tand21}
\end{figure}

As shown in Fig.~\ref{Fig:alpham1}, we get restrictions on the absolute masses
of the neutrinos by combining mixing results from solar neutrino observations
and bounds on $|\langle m\rangle|$ from $0\nu\beta\beta$-experiments.  For a
given $|\langle m\rangle|$ and given values of the mixing angles, the upper
bound is, for degenerated masses:
\begin{equation}\label{maxm1}
\max(m_1) = \frac{|\langle m\rangle|}{2U_{e1}^2-1}, \qquad
\mbox{if $\tan^2\theta_{12} < 1$}.
\end{equation}
For example, with $U_{e3}=0$ and $|\langle m\rangle|=0.26 ~\mbox{eV}$, the
best-fit angle $\tan^2\theta_{12}=0.3$ (see Fig.~\ref{Fig:tand21}) gives
$\max({m_1}) \simeq 0.5~\mbox{eV}$ and $\tan^2\theta_{12}=0.8$ gives
$\max({m_1}) \simeq 2.3 ~\mbox{eV}$.  For the 90\% C.L. bound in
Eq.~(\ref{bounds}), these mixings give respectively
$\max(m_1)\simeq0.65~\mbox{eV}$ and $\max(m_1)\simeq 3.0~\mbox{eV}$.  (For
degenerated masses, the lowest possible mass corresponds to $m_1 = |\langle
m\rangle|$.)

If a future positive $0\nu\beta\beta$-signal should allow the expression in
Eq.~(\ref{maxm1}) to be larger than the observed mass bound, we get
restrictions on the allowed mixing and phases, as illustrated in
Fig.~\ref{Fig:alpham1}.  If we assume Spectrum 1 and two generations, the
relation between the remaining neutrino parameters can be written (from
Eq.~(\ref{spect1}))
\begin{equation}
\cos \alpha_1 = \frac{|\langle m\rangle|^2-m_1^2(1-2U_{e1}^2U_{e2}^2)}
{2m_1^2U_{e1}^2U_{e2}^2}, \qquad \mbox{if $\Delta m_{21}^2 \ll m_1^2$}.
\end{equation}

In the upper part of Fig.~\ref{Fig:tand21} we compare the most
probable solution for the solar neutrino problem, the LMA region, with
different hypothetical contours for $|\langle m\rangle|$ under the
assumption of $m_1 =0$.  Whether or not the LMA region is the
correct one, is likely to be determined in the next few years by the KamLAND
experiment \cite{Piepke:2001tg}.
The LOW and VO regions are not included in this part because the
corresponding $|\langle m\rangle|$-values are far below the values which can be
detected in the foreseeable future.
The Majorana phases have been chosen to
give the smallest possible $|\langle m\rangle|$-value, which in our examples
implies $\alpha_1 = \alpha_2 =\pi$.  The $|\langle m\rangle|$-values
indicated in this hierarchical case ($m_1=0$) are at the border of the
sensitivity of the most optimistic GENIUS proposal, see Table
\ref{Tab:masses}.

For the near-degenerated case $m_1=0.1 ~\mbox{eV}$, the $|\langle
m\rangle|$-values are compared to the two most favoured solar neutrino
solutions, the LMA and LOW regions.  For clarity we don't indicate the VO
region; its allowed region of mixing largely overlaps with that of the LMA
and LOW, and like the last one it includes $\tan^2\theta_{12}=1$, i.e., it
allows very large masses.  The LOW region extends to lower values of the
mass-squared difference than that shown in the figure,  but this part is
covered by the shown range of mixing, and the $|\langle m\rangle|$-values
have very weak dependence on such small mass-squared differences.  For other
phases or higher masses than those considered in the lower part of
Fig.~\ref{Fig:tand21}, the near-vertical contours will be shifted towards
larger mixing.  In other words, if, for example, the effective Majorana mass
should turn out to be $|\langle m\rangle|= 0.05~\mbox{eV}$, the region to the
left of the corresponding contours would require $m_1 < 0.1 ~\mbox{eV}$ or
$U_{e3} > 0.1$.  For that particular $|\langle m\rangle|$-value, we showed in
Fig.~\ref{Fig:alpham1} the range in $m_1$ as a function of the phase
$\alpha_1$ for four different mixings.  To scale the contours in the lower
part of Fig.~\ref{Fig:tand21} for larger $m_1$-values, we can use the
relation $|\langle m\rangle| \propto m_1$, which in this case is valid
because Spectrum 1 and small $U_{e3}$ require a weaker constraint than the
general one, which is
\begin{equation}\label{prop}
|\langle m\rangle| \propto m_1, \qquad
\mbox{if $\Delta m_{\rm atm}^2 \ll m_1^2$}.
\end{equation}
For negligible $U_{e3}$, this inequality should read $\Delta m_{\odot}^2 \ll
m_1^2$.  From Eq.~(\ref{prop}) and the lower part of Fig.~\ref{Fig:tand21} we
see that the LMA region, if it is confirmed, should lead to a positive signal
in GENIUS if we have Spectrum 1, Majorana neutrinos and $m_1 \gtrsim
0.1~\mbox{eV}$.  For LOW there is a slight chance of getting positive results
for $m_1=0.1~\mbox{eV}$, this probability increases as the masses get bigger.

\begin{figure}[t]
\begin{center}{\mbox{\epsffile{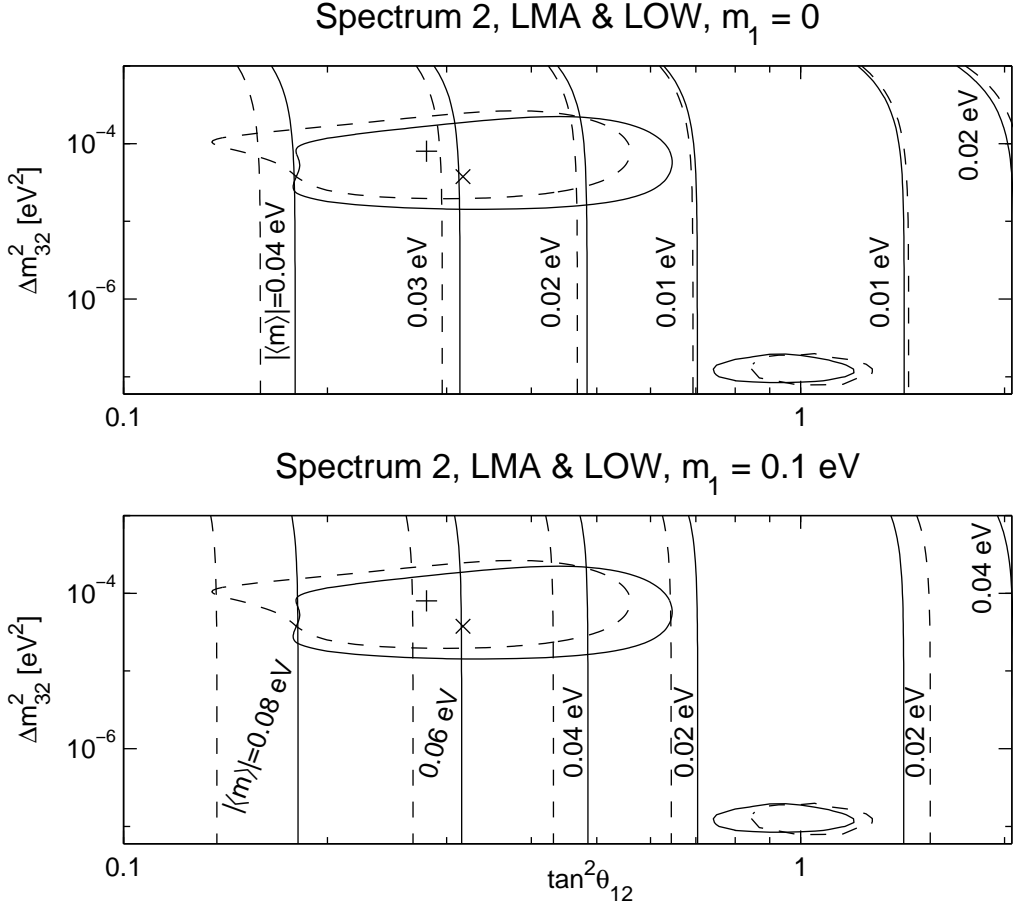}}}
\end{center}
\vspace*{-8mm}
\caption{Same as Fig.~\ref{Fig:tand21}, but for Spectrum 2.  The dashed lines
and ``+'' are for $U_{e3}=0.2$.}
\label{Fig:tand32}
\end{figure}
When $|\langle m\rangle| = 0.01 ~\mbox{eV}$ is assumed as the future
threshold for a positive $0\nu\beta\beta$-signal, we see from the upper part
of Fig.~\ref{Fig:tand32} that for Spectrum 2 the whole LMA region (95\% C.L.)
can be covered by reachable $|\langle m\rangle|$-values, even for $m_1=0$.
The width of the allowed solar neutrino regions depends somewhat on how the
neutrino data are treated (only rates or also spectral information, errors of
cross sections, etc.). Thus the LMA region could extend below the future
lower bound for $0\nu\beta\beta$-observations.  Spectrum 2 would have to be
discarded if the LMA region is confirmed and if an $|\langle m\rangle|$-value
will be found lying to the right of the new allowed LMA contour in the upper
part of Fig.~\ref{Fig:tand32}.  For Spectrum 2 the proportionality relation
in Eq.~(\ref{prop}) is not valid unless $m_1 \gtrsim 0.4 ~\mbox{eV}$. Above
this value the two spectra are quite similar.

It should be noted that the closed contours in Figs.~\ref{Fig:tand21} and
\ref{Fig:tand32} are based only on the total rates measured in solar neutrino
detectors.  As shown in \cite{Fogli:2001xt}, when CHOOZ data are included in
these calculations, the allowed regions decrease as $U_{e3}$ increases.

\begin{figure}[th]
\begin{center}{\mbox{\epsffile{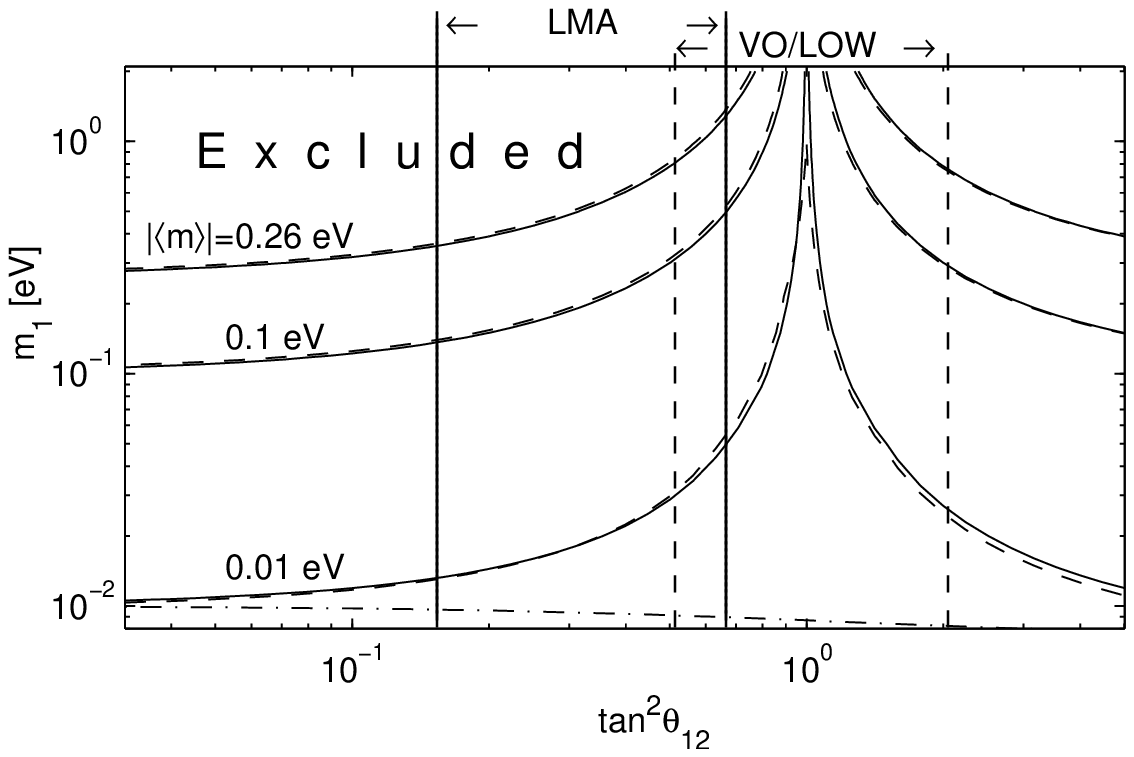}}}
\end{center}
\vspace*{-8mm}
\caption{Maximal allowed values of $m_1$ for Spectrum 1.  Solid curves:
$|\langle m\rangle|= 0.26 ~\mbox{eV}$, 0.1~\mbox{eV} and 0.01~\mbox{eV}, all
for $U_{e3}=0$; dashed curves: $U_{e3}=0.1$.  Spectrum 2 would have produced
practically the same results, except for $|\langle m\rangle|=0.01~\mbox{eV}$
when $m_1 \lesssim 0.1 ~\mbox{eV}$.  Dash-dotted curve: {\it minimum} $m_1$
for $|\langle m\rangle| = 0.01~\mbox{eV}$ and Spectrum 1.  The regions
between the vertical pairs of lines are the allowed LMA and VO/LOW
solutions.}
\label{Fig:tanm1}
\end{figure}

In Fig.~\ref{Fig:tanm1} we show how the largest allowed $m_1$-value for
Spectrum 1 changes as a function of $\tan^2\theta_{12}$ over a region
covering the 95\% C.L. LMA and VO/LOW fits for $|\langle m\rangle|=0.26
~\mbox{eV}$, 0.1 eV and 0.01 eV.  (As noted above, the VO and LOW regions are
not shown separately because they overlap to some extent and both of them
cover the maximal mixing case.)  The $m_1$-values for the leftmost part of
the curves are practically the same as those for the SMA region. The variation
of $|\langle m\rangle|$ within the SMA region is totally negligible.  The
highest $m_1$-value in Fig.~\ref{Fig:tanm1} is close to the cosmological
bound, see Table \ref{Tab:masses}.  If the allowed mass value for the Z-burst
theory is to be confirmed and/or sharpened, we see from this figure that in
case of Majorana neutrinos, an observation of $0\nu\beta\beta$ seems very
likely, especially if the LMA region should be confirmed by KamLAND.

\begin{figure}[ht]
\begin{center}{\mbox{\epsffile{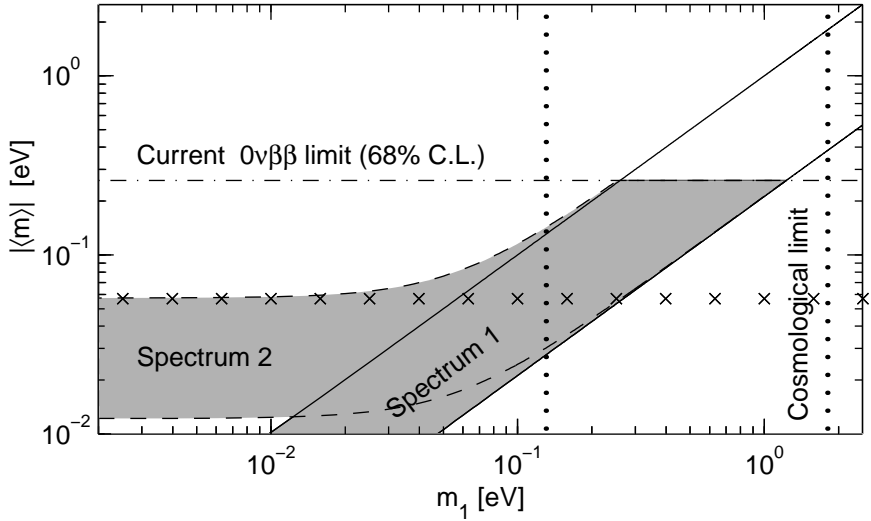}}}
\end{center}
\caption{Allowed bands in the LMA region (gray shaded). The solid lines
represent the borders for Spectrum 1, the dashed curves give the allowed
region for Spectrum 2. We have shown today's upper limit, $|\langle
m\rangle|=0.26 ~\mbox{eV}$, on neutrinoless double beta decay and the
cosmological bound, (1/3)($\sum_j m_j \lesssim 5.5 ~\mbox{eV}$).  The lowest
horizontal axis and the dotted vertical line show the technical sensitivity
of planned experiments for measuring these two kinds of phenomena
\cite{Croft:1999mm}.  The line of crosses represents the value of
$\sqrt{\Delta m_{\rm atm}^2}$.}
\label{Fig:Mee}
\end{figure}

It is instructive to compare the data from $0\nu\beta\beta$ and solar
neutrino measurements to determine the allowed regions in a plane spanned by
$m_1$ and $|\langle m\rangle|$.  This is shown in Fig.~\ref{Fig:Mee}, where
the allowed regions (shaded) are determined by the LMA fit and bounded by the
current limit from neutrinoless double beta decay.  The highest and lowest
$|\langle m\rangle|$-values are for a given $m_1$ found with $\alpha_1=0$ and
$\pi$, respectively.  For the atmospheric mass-squared difference we used the
best-fit value $\Delta m_{\rm atm}^2 = 3.3 \times 10^{-3}~\mbox{eV}^2$.
(Since the shaded region changes insignificantly
within the allowed $U_{e3}$ range, the figure is drawn for $U_{e3}=0$.)
Again, we see that Spectrum 1, as opposed to Spectrum 2, allows for lower
$|\langle m\rangle|$-values than those measurable in the planned experiments.
This figure also illustrates the confluence of the two spectra for
increasing masses.

If there should be {\em no} signal above the claimed future sensitivity,
i.e., $|\langle m\rangle| < 0.01 ~\mbox{eV}$, then we have one or two of the
possibilities:
\begin{itemize}
\item The neutrinos are of Dirac character.
\item Spectrum 1 and $m_1$ below the $|\langle m\rangle|=0.01~\mbox{eV}$
contour in Fig.~\ref{Fig:tanm1}.
\item Spectrum 2 and mixing close to maximal.
\end{itemize}
For Spectrum 2 the SMA and the 95\% C.L. part of the LMA region would be
discarded.
\section{Summary}
We have discussed the interrelation between solar neutrino data and current
and future results from both neutrinoless double beta decay experiments and
cosmological observations.  It is qualitatively shown how a Majorana phase
and a mixing angle could be related after eventual future measurements of the
effective Majorana mass, $|\langle m\rangle|$ and the neutrino masses, $m_j$.
Values of $|\langle m\rangle|$ are compared to the LMA and VO/LOW solutions
in terms of $m_1$.  It is also shown that the allowed non-zero
$U_{e3}$-values hardly affect the conclusion.  The two spectra are
indistinguishable for $m_1 \gtrsim 0.4~\mbox{eV}$.  The four main solutions
to the solar neutrino problem can be related to the
$0\nu\beta\beta$-phenomenology as follows:
\begin{itemize}
\item{SMA:} Due to the smallness of $U_{e2}$ and $U_{e3}$, the $|\langle
m\rangle|$ expressions are proportional to $m_1$ (Spectrum 1) or to $m_2$
(Spectrum 2), and they have a very weak dependence on the Majorana phases.
The current bound ($|\langle m\rangle| < 0.26~\mbox{eV}$) leads to the mass
bound $m_1 \lesssim 0.3~\mbox{eV}$, which excludes a high degree of
degeneracy.  The Spectrum 2 part is, for any $m_1$-value, within the
sensitivity of GENIUS.

\item{LMA:} Due to the involved mixing, the $|\langle m\rangle|$-value is
quite dependent on one of the Majorana phases. The neutrino masses are
bounded by $m_1 \lesssim 1.5~\mbox{eV}$ ($U_{e3}=0$), and a high degree of
degeneracy is allowed.  For planned experiments, Spectrum 1 is perhaps below
observation for truly hierarchical masses.  The entire allowed region for
Spectrum 2 is within the GENIUS sensitivity, but just barely for its
lowest $|\langle m\rangle|$-values.  If a bound $|\langle m\rangle|
\lesssim 0.01~\mbox{eV}$ should be established, then Spectrum 2 would be
seriously disfavoured.

\item{LOW:} Not detectable for a normal hierarchy unless $U_{e3} \gtrsim
0.2$.  This region allows $U_{e1} = U_{e2}$, which implies $|\langle
m\rangle| \simeq 0$ for very large masses.

\item{VO:} Similar to the LOW region.
\end{itemize}

\end{document}